\documentclass[a4paper,11pt]{article}
\pdfoutput=1
\usepackage{jheppub}

\usepackage[utf8]{inputenc}
\usepackage{subfigure} 
\usepackage{tikz}
\usetikzlibrary{backgrounds,automata}
\newcommand{\be}{\begin{equation}}
\newcommand{\ee}{\end{equation}}
\newcommand{\bea}{\begin{equation}\begin{aligned}} 
\newcommand{\eea}{\end{aligned}\end{equation}}
\newcommand{\MeV}{{\rm MeV}}
\newcommand{\TeV}{{\rm TeV}}
\newcommand{\GeV}{{\rm GeV}}
\newcommand{\mpl}{M_P}
\newcommand{\td}{\mathrm{d}}

\newcommand{\rR}{\rho_{R}}
\newcommand{\sv}{\langle\sigma v\rangle}
\newcommand{\Trh}{T_\text{RH}}
\newcommand{\Tmax}{T_\text{max}}
\newcommand{\gs}{g_\star}
\newcommand{\gss}{g_{\star s}}
\newcommand{\Br}{\text{Br}_\text{DM}}
\newcommand{\bigO}{\mathcal O}

\def\SSlash#1{\hskip 0.1 cm \slash\hskip -0.3 cm #1}

\title{UV Freeze-in in Starobinsky Inflation}

\author[a]{Nicolás Bernal,}
\author[b,c]{Javier Rubio}
\author[d]{and Hardi Veerm\"ae}

\affiliation[a]{Centro de Investigaciones, Universidad Antonio Nariño\\
Carrera 3 Este \# 47A-15, Bogotá, Colombia}
\affiliation[b]{Department of Physics and Helsinki Institute of Physics\\
PL 64, FI-00014 University of Helsinki, Finland}
\affiliation[c]{Centro de Astrof{\'i}sica e Gravita\c c\~ao-CENTRA, Departamento de F{\'i}sica,
Instituto Superior T{\'e}cnico-IST, Universidade de Lisboa-UL,
Avenida Rovisco Pais 1, 1049-001, Lisboa, Portugal}
\affiliation[d]{National Institute of Chemical Physics and Biophysics\\
R\"avala 10, 10143 Tallinn, Estonia}

\emailAdd{nicolas.bernal@uan.edu.co}
\emailAdd{javier.rubio@helsinki.fi}
\emailAdd{hardi.veermae@cern.ch}

\abstract{In the Starobinsky model of inflation, the observed dark matter abundance can be produced from the direct decay of the inflaton field only in a very narrow spectrum of close-to-conformal scalar fields and spinors of mass $\sim 10^7$ GeV. This spectrum can be, however, significantly broadened in the presence of effective non-renormalizable interactions between the dark and the visible sectors. In particular, we show that UV freeze-in can efficiently generate the right dark matter abundance for a large range of masses spanning from the keV to the PeV scale and arbitrary spin, without significantly altering the heating dynamics. We also consider the contribution of effective interactions to the inflaton decay into dark matter.
\newpage}

\begin{document}
\begin{flushright}
    PI/UAN-2020-673FT\\
    HIP-2020-17/TH
\end{flushright}

\maketitle
\flushbottom

\section{Introduction} \label{sec:in}

The celebrated Starobinsky model of inflation, in which cosmological perturbations were originally computed~\cite{Mukhanov:1981xt}, remains in excellent agreement with observations~\cite{Aghanim:2018eyx, Akrami:2018odb} even four decades later of being proposed. This scenario is based on the presence of a $R^2$ term on top of the Einstein-Hilbert action and leads generically to three distinct epochs: $i)$ an inflationary state at large curvature values, $ii)$ a matter-dominated era following inflation and $iii)$ a crossover regime where particle production takes place and the Universe transitions from matter domination to radiation domination.  

One of the attractive features of Starobinsky inflation is that gravitational degrees of freedom are responsible for both   inflation and the onset of the hot big bang. Compared to other inflationary scenarios, the Starobinsky model is highly predictive -- it contains a single free parameter which is determined by the observed amplitude of primordial density fluctuations~\cite{Martin:2013tda}. All interactions with matter fields in its dual scalar-tensor representation are universally dictated from their conformal properties, reducing the usual uncertainty in post-inflationary model building. 

The heating stage in Starobinsky inflation proceeds through the gravitational particle production of non-conformally coupled fields. This slow but continuous process produces a subdominant radiation component even before the heating of the  Universe is complete.\footnote{The time of (complete) heating is usually defined as the moment at which the energy densities of radiation and the inflaton become equal. The heating process is often referred to as ``reheating".} During this period, the temperature of the radiation bath exceeds the heating temperature (see e.g. Refs.~\cite{McDonald:1989jd, Chung:1998rq, Giudice:2000ex, Allahverdi:2002nb, Allahverdi:2002pu}), opening the door to the indirect production of dark matter (DM) out of the Standard Model (SM) plasma.

In this paper, we study the direct and indirect production of DM in the Starobinsky model of inflation with a particular focus on the so-called UV freeze-in mechanism~\cite{Hall:2009bx, Elahi:2014fsa, Bernal:2017kxu}. This mechanism is highly sensitive to the dynamics of the thermal bath during the intermediate period between inflation and radiation domination~\cite{Garcia:2017tuj, Bernal:2019mhf}, and therefore, it is natural to be analysed within a specific heating setup like the one under consideration (for related studies, see e.g. Refs.~\cite{Chen:2017kvz, Bernal:2018qlk, Bhattacharyya:2018evo, Chowdhury:2018tzw, Kaneta:2019zgw, Banerjee:2019asa, Dutra:2019xet, Dutra:2019nhh, Mahanta:2019sfo, Cosme:2020mck, Bernal:2019mhf, Garcia:2020eof, Bernal:2020bfj}).
Unlike IR freeze-in scenarios, where the bulk of the DM abundance is produced at low scales, typically when the temperature of the SM thermal bath is of the order of the DM or the mediator mass, UV freeze-in is mostly effective at high temperatures. In particular, this mechanism is possible if the thermally averaged cross section connecting the dark and visible sectors grows with temperature. The last feature implies that the two sectors are effectively connected by higher dimensional operators, which can naturally appear in the presence of heavy mediators once these are integrated out (see e.g. Ref.~\cite{Bernal:2019mhf}). As such operators are not conformally invariant, they will inevitably couple both sectors to the inflaton field, inducing additional decay channels. 

We consider a minimalistic matter sector containing the SM and an additional DM candidate. After showing that the direct perturbative production of DM via two-body inflaton decays is only viable for close-to-conformal scalar fields or spinors of very specific mass, we demonstrate that the observed DM density can be produced via the above UV freeze-in mechanism. In fact, if the visible and DM sectors are connected by higher dimensional operators, the whole DM relic density can be viably produced by UV freeze-in for a \textit{broad range of DM masses} between the keV and the PeV scale, and \textit{arbitrary spin}. These new interactions can leave the heating dynamics in the Starobinsky model unaltered. However, the higher dimensional operators open new decay channels into the dark sector which, in general, cannot be a priori neglected as they may compete with the UV freeze-in process.

This paper is organized as follows.  In Section~\ref{sec:Starobinsky} we review the Starobinsky model of inflation, connecting it to its dual scalar theory. The heating dynamics is studied in Section~\ref{sec:reheat}. The production of the DM relic abundance via the UV freeze-in mechanism is discussed in Section~\ref{sec:discussion}.  Finally, we provide a summary and some concluding remarks in Section~\ref{sec:end}. Throughout the paper we use the metric signature $(-,\,+,\,+,\,+)$ and natural units $\hbar = c = 1$.

\section{The Starobinsky Model} \label{sec:Starobinsky}

Consider the action\footnote{We include a tilde to denote Jordan-frame fields in order to have simpler expressions in the scalaron-frame~\eqref{eq:EFaction}, where most of the computations in this paper will be performed.}~\cite{Starobinsky:1980te,Starobinsky:1981vz,Starobinsky:1983zz,Kofman:1985aw}
\be\label{eq:S_staro}
	S = \frac{\mpl^2}{2}\int \td^4x\sqrt{-\tilde g}\left[\tilde R + \frac{\tilde R^2}{6M^2}\right] + S_{M}(\tilde \varphi,\tilde \psi,\tilde A_\mu)\,,
\ee
with $M_P=(8\pi\,G)^{-1/2}\simeq2.4\times 10^{18}$~GeV the reduced Planck mass, $M$ a mass parameter to be determined from observations and $S_M$ a matter action accounting for the SM model content and other potential beyond the SM sectors including DM. For the sake of concreteness, we will assume this matter action to include $n_\varphi$ scalar fields ($\tilde\varphi$) non-minimally coupled to gravity, $n_\psi$ fermions ($\tilde \psi$) and $n_{A}$ vector fields ($\tilde A_\mu$), namely $S_M=S_0+S_{1/2}+S_1$ with
\bea\label{eq:JF} 
    S_0 &= \int \td^4 x \sqrt{-\tilde g}\sum_{n_\varphi}\left[ 
    - \tilde  D_\mu\tilde \varphi^*\tilde D^\mu\tilde \varphi  - (m^2_{\varphi}-\xi_\varphi \tilde R ) |\tilde \varphi|^2 -\lambda |\tilde \varphi|^4 \right]\,,  
    \\
    S_{1/2} &= \int \td^4 x \sqrt{-\tilde g} \sum_{n_\psi}\left[-i\bar{\tilde \psi} \SSlash{D} \tilde \psi-m_\psi \bar{\tilde\psi}\tilde \psi \right]\,, 
    \\
    S_1 &= \int \td^4 x \sqrt{-\tilde g} \sum_{n_A}\left[-\frac14 \tilde F^{\mu\nu}\tilde F_{\mu\nu} 
\right]\,.
\eea
Here $\tilde F^{\mu\nu}$ stands for the field strength of $\tilde A_{\mu}$ , $\tilde D_{\mu}\,\tilde \varphi = (\partial_{\mu} - i g \tilde A_{\mu})\tilde \varphi$ and $\tilde{\SSlash{D}}\tilde \psi =\tilde e^{\mu}{}_\alpha\gamma^\alpha(\partial_\mu-{\tilde \Gamma}_\mu-ig\tilde {A}_\mu)\tilde \psi$ denote respectively scalar and spinor covariant derivatives with $g$ a gauge coupling constant, $\tilde e^{\mu}{}_{\alpha}$ a vierbein field and  $\tilde \Gamma_\mu\equiv -\frac12\Sigma^{\alpha\beta}\tilde e^\lambda{}_{\alpha}\tilde \nabla_\mu \tilde e_{\lambda\beta}$ a spin connection, with $\Sigma^{\alpha\beta}=-\Sigma^{\beta\alpha}=\frac14[\gamma^\alpha,\gamma^\beta]$ the Lorentz group generators~\cite{Birrell:1982ix}. Space-time indices are raised and lowered with the $\tilde g_{\mu\nu}$ metric.

\begin{figure}[t!]
	\begin{center}
		\includegraphics[height=0.34\textwidth]{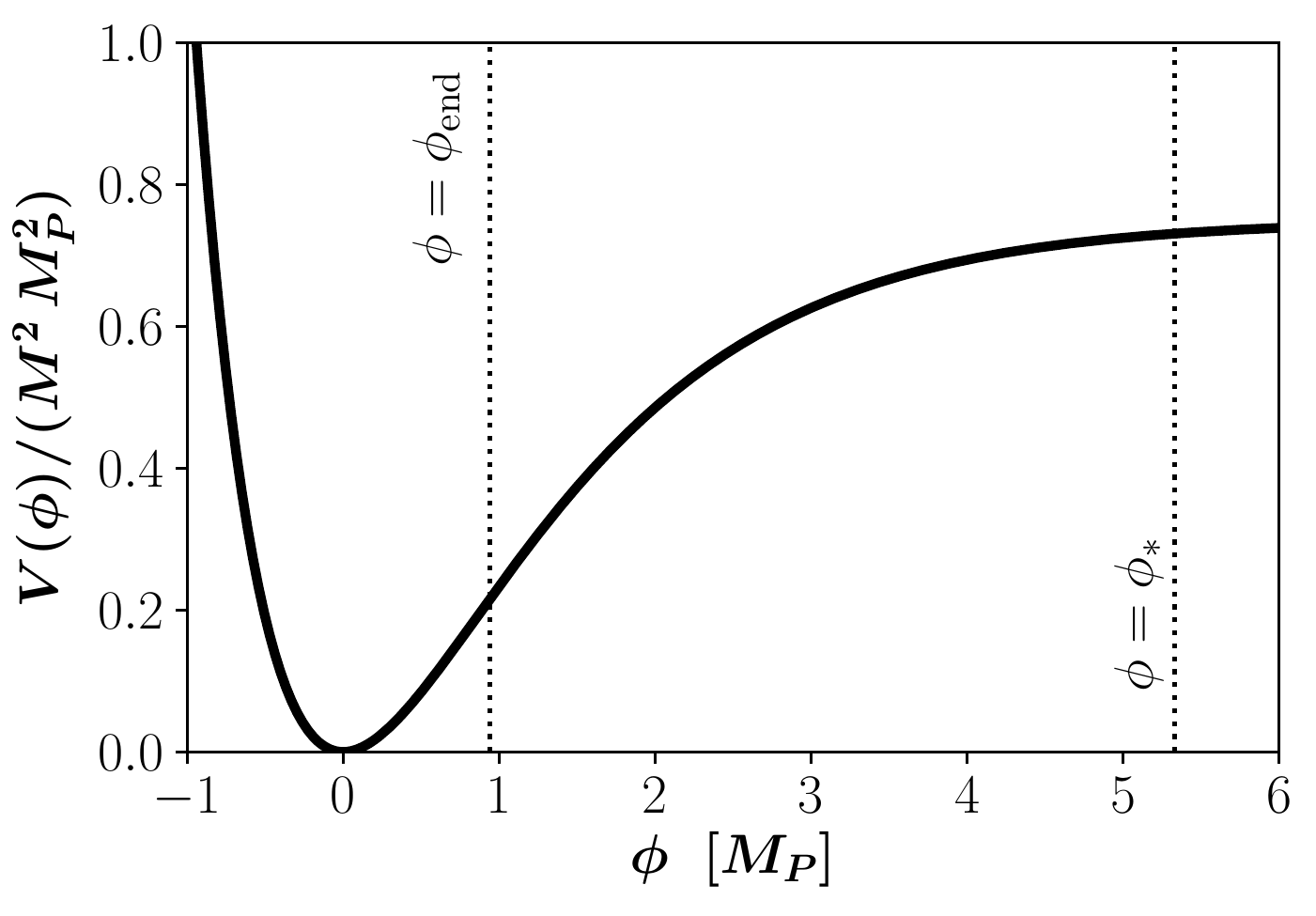}
		\includegraphics[height=0.34\textwidth]{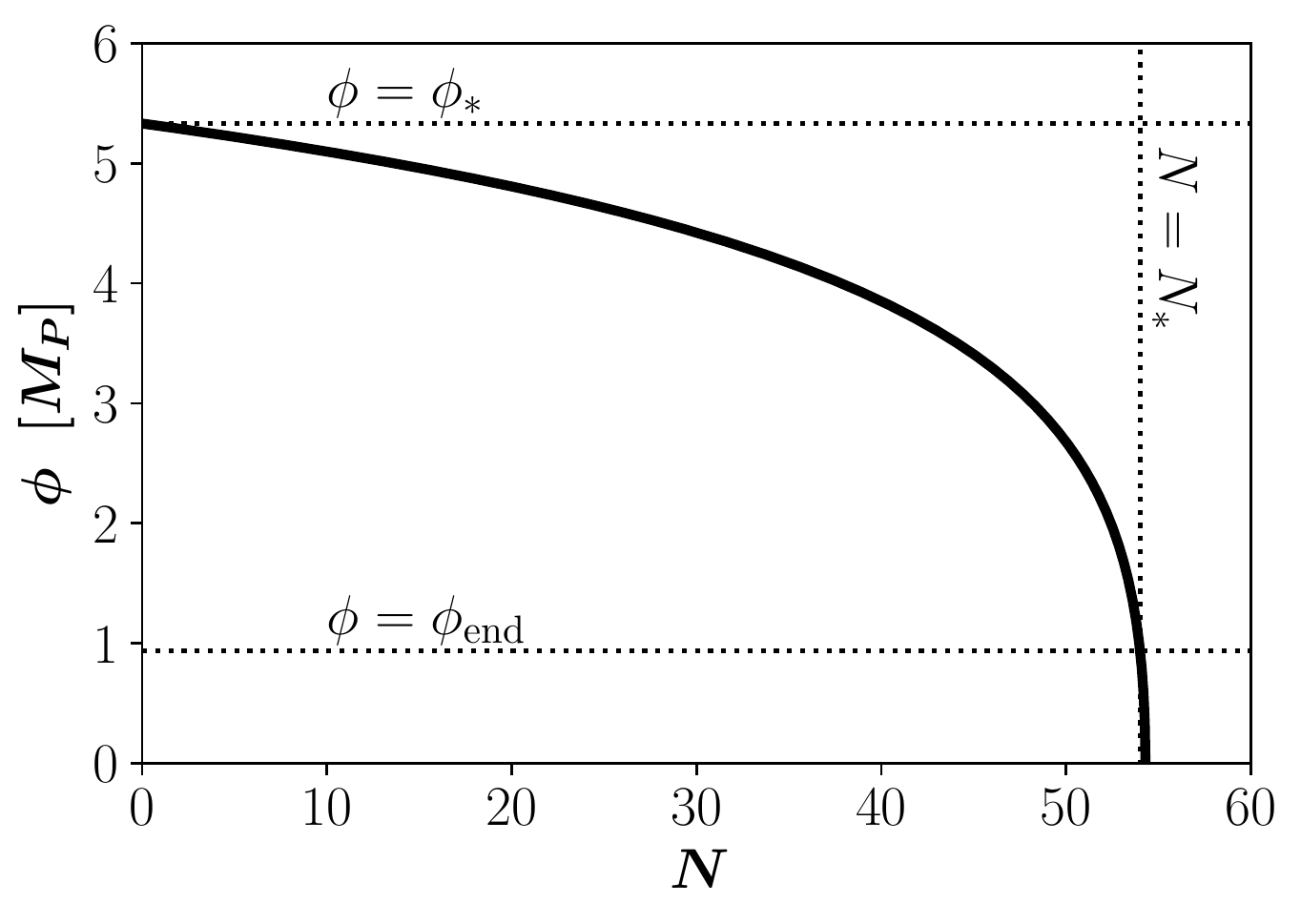}
		\caption{Potential for the Starobinsky model of inflation in the scalaron-frame (left panel) and evolution of the scalaron field as a function of the number of number of $e$-folds (right panel). As a reference, dotted lines corresponding to $\phi=\phi_*$, $\phi=\phi_\text{end}$ and $N=N_*=54$ are overlaid (cf.~Section~\ref{sec:reheat}).}
		\label{fig:Starobinsky}
	\end{center}
\end{figure}

The presence of the $\tilde R^2$ term in Eq.~\eqref{eq:S_staro} allows for an inflationary state able to generate the observed amount of  primordial density perturbations for suitable values of the mass parameter $M$~\cite{Starobinsky:1980te,Starobinsky:1981vz}. This can be seen explicitly by rewriting $R^2$ in terms of an auxiliary field $\Phi$,\footnote{Note that upon integrating out $\Phi$, one gets back the original theory.}
\be
	S = \int \td^4x\sqrt{-\tilde g}\left[\frac{1}{2}(\mpl^2 + \xi \,\Phi^2)\tilde R - \frac{\lambda}{4} \Phi^4\right]+ S_{M}(\tilde \varphi,\tilde \psi,\tilde A_\mu)\,, \hspace{8mm} \lambda \equiv  \frac{3M^2 \xi^2}{2\mpl^2}\,.
\ee
Now, by performing a Weyl transformation $\tilde g_{\mu\nu} = \Omega^2\,g_{\mu\nu}$ with the conformal factor 
\be\label{eq:Omega}
    \Omega^2=  1 + \frac{\xi\,\Phi^2}{\mpl^2}=\exp\left(\sqrt{\frac23}\frac{\phi}{\mpl}\right)\,,
\ee
the action~\eqref{eq:S_staro} is recast as~\cite{Stelle:1977ry, Whitt:1984pd, Mukhanov:1989rq}
\be\label{eq:EFaction}
	S =  \int \td^4x\sqrt{-g}\left[\frac{\mpl^2}{2} R - \frac{1}{2}g^{\mu\nu} \partial_{\mu}\phi\, \partial_{\nu}\phi  -  V(\phi)\right]+ S_{M}( \varphi,\psi, A_\mu)\,,
\ee
where 
\be\label{eq:V}
	V(\phi) = \frac34 \mpl^2\,M^2\left[1-e^{-\sqrt{\frac23}\frac{\phi}{\mpl}}\right]^2
\ee
is the potential of the so-called \textit{scalaron} field $\phi$ (cf.~Fig.~\ref{fig:Starobinsky}). In terms of the conformally rescaled fields,
\be\label{eq:conf_fields}
    \varphi = \Omega^{-1}\tilde \varphi\,,\hspace{7mm}
    \psi = \Omega^{-3/2}\tilde \psi\,, \hspace{7mm}
    {A}_{\mu} = \tilde A_{\mu}\,, \hspace{7mm}
    e^{\mu}{}_\alpha=\Omega^{-1} \tilde e^{\mu}{}_\alpha\,, \hspace{7mm} {\Gamma}_{\mu}=\tilde \Gamma_{\mu}\,,
\ee
the matter action $S_M$ reads
\bea\label{SM_EF}
    S_0  = \int \td^4 x \sqrt{-g}&\sum_{n_\varphi}\bigg[
  - (D_\mu\varphi)^*D^\mu\varphi- (\Omega^{-2} \, m_\varphi^2- \xi_\varphi R)|\varphi|^2 \\ & 
  +  (1 + 6\xi_\varphi) |\varphi|^2 \left(\square \ln \Omega - (\partial \ln \Omega)^2 \right) -\lambda |\varphi|^4  \bigg]\,, \\ 
    S_{1/2}  = \int \td^4 x \sqrt{-g}&\sum_{n_\psi}\left[-i\bar\psi\SSlash{D}\psi+\Omega^{-1} m_\psi \bar\psi\psi \right]\,, \\
    S_1  = \int \td^4 x \sqrt{-g} &\sum_{n_A}\left[-\frac14 F^{\mu\nu}F_{\mu\nu} 
\right]\,,
\eea
with $F^{\mu\nu}$ the field strength of $A^{\mu}$ and $D^{\mu}$, $\SSlash{D}$ scalar and spinor covariant derivatives depending on the corresponding untilded quantities. The kinetic terms of fermions and vectors as well as contact interaction terms of mass dimension 4 are Weyl invariant. Note, however, that this does not apply to dimensionful parameters, explicit mass terms or the kinetic terms of non-conformally coupled scalar fields such as the SM Higgs. Since the matter sector still contains non-minimal couplings to gravity, we will refer to this frame as the \textit{scalaron frame}, refraining from using the more common \textit{Einstein-frame} terminology.

The approximate shift-symmetry of Eq.~\eqref{eq:EFaction} at large field values can be understood as the scalaron-frame manifestation of  the emergent scale symmetry of Eq.~\eqref{eq:S_staro} in the large curvature regime. In this form, the usual slow-roll techniques allow to compute the amplitude $\cal P$ of the power spectrum of primordial density fluctuations, its tilt $n_s$ and the tensor to scalar ratio $r$~\cite{Mukhanov:1981xt, Starobinsky:1983zz, Mukhanov:1989rq}, 
\be\label{eq:predict}
    \mathcal{P}=\frac{N_*^2}{24\pi^2}\left(\frac{M}{M_P}\right)^2\,,
    \hspace{12mm}	n_s\simeq  1-\frac{2}{N_*}\,, 
    \hspace{12mm}	r \simeq\frac{12}{N_*^2}\,. 
\ee
The scalaron mass $M$ is determined by the COBE normalization ${\cal P}=2.1\times 10^{-9}$~\cite{Akrami:2018odb},
\be\label{Mvalue}
    M\simeq 1.3\times  10^{-5} \left(\frac{54}{N_*}\right) M_P \,.
\ee
The precise number of $e$-folds $N_*$ to be inserted in these observables depends on the reheating temperature $T_{\rm RH}$ at which the Universe becomes dominated by radiation, namely~\cite{Liddle:2000cg}
\be\label{eq:N}
    N_*\simeq  54+ \frac{1}{3} \log \left(\frac{T_{\rm RH}}{10^{9}\,{\rm GeV}}\right)\,.
\ee
This temperature depends itself on the details of the heating stage, which we now proceed to describe.

\section{Standard Model Production}\label{sec:reheat}

The heating stage in the Starobinsky model of inflation proceeds through the gravitational particle production of non-conformally coupled fields.\footnote{The formation of massive oscillons, typically associated with potentials that away from the minimum are shallower than quadratic~\cite{Amin:2011hj}, does not take place in this setting~\cite{Takeda:2014qma}.} It can be computed via the Bogoliubov's method in the Jordan frame~\eqref{eq:S_staro}~\cite{Starobinsky:1980te,Starobinsky:1981vz} or using standard perturbative techniques in the scalaron frame~\cite{Watanabe:2006ku,Watanabe:2010vy}, getting the same result in both cases~\cite{Rudenok:2014daa,Ema:2016hlw}. We will follow here the second approach.

Taylor expanding the conformal factors in Eq.~\eqref{SM_EF} in $\phi/\mpl$, we obtain a series of trilinear interactions involving the scalaron field $\phi$. Since the only non-conformally coupled field in the SM is the Higgs boson, the scalaron will only decay to it at tree level. The corresponding two-body decay width is given by 
\be\label{Gamma_varphi}
    \Gamma_{\rm tree,\, SM} \simeq
    \frac{1}{24\pi}\left[1+6\,\xi_H+2\frac{m_H^2}{M^2}\right]^2\frac{M^3}{M_P^2}\,,
\ee
where we have taken into account that the mass of the Higgs is much smaller than the scalaron mass, $m_H \ll M$, neglecting therefore phase-space suppression factors. We remark that the non-minimal coupling $\xi_H$ affects the Higgs potential and thus the stability of the electroweak vacuum~\cite{Fumagalli:2019ohr}. In particular, negative but moderate values\footnote{The precise range depends on the value of top pole mass~\cite{Figueroa:2017slm}.} of $\xi_H$ enhance the effective Higgs mass during inflation~\cite{Espinosa:2007qp,Herranen:2014cua} without generating strong tachyonic instabilities during the heating stage~\cite{Herranen:2014cua,Kohri:2016wof,Ema:2016kpf,Figueroa:2017slm}. Having this in mind, we will restrict ourselves to a conservative range $-1/6\leq \xi_H\leq 0$ in what follows.  Note also that, at the leading-order, the decay width $\Gamma_{\rm tree,\, SM}$ vanishes when the non-minimal coupling of the Higgs field to gravity is close to its conformal value $\xi_{H}=-1/6$.

\begin{figure}[t!]
	\begin{center}
		\includegraphics[width=0.3\textwidth]{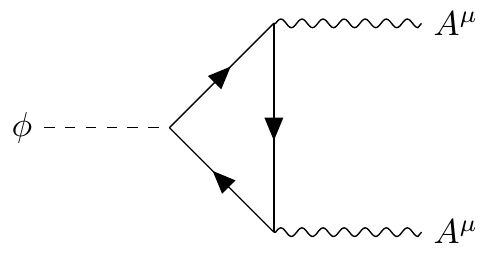}
		\includegraphics[width=0.3\textwidth]{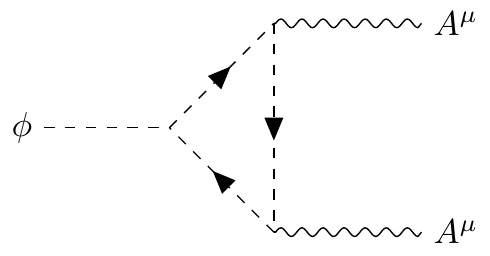}
         \includegraphics[width=0.3\textwidth]{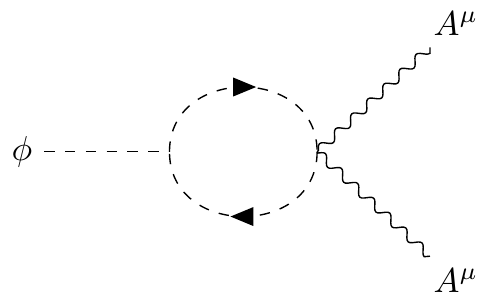}
  \caption{Feynman diagrams for the 1-loop (anomalous) scalaron decay into gauge bosons.}\label{fig:1loop}
	\end{center}
\end{figure}

Although absent at tree level, the anomalous decay of the scalaron field into gauge bosons is possible at 1-loop (cf.~Fig.~\ref{fig:1loop}). In particular, the breaking of scale symmetry during the regularization process translates into an induced breaking of the gauge conformal symmetry. If the intermediate states running in the loops are integrated out, the effect of the anomaly can be expressed through a \textit{local effective action}
\be\label{eq:anom}
    S_a=\int \td^4 x \sqrt{-g}\left[\frac{\beta_h(g)}{2g}(\ln{\Omega^2}){F}^{\mu\nu}{F}_{\mu\nu}\right]\,,
\ee
with
\be\label{beta}
    \beta_h(g)=-\frac{g^3 \, b}{(4\pi)^2} \,,\hspace{10mm} 
    b=\sum_\text{internal} \frac{1}{3}\left[\left(11 N-1\right)N_{1}- N_{0} -2 N_{1/2}\right]\,,
\ee
the contribution of the internal fields to the 1-loop Callan-Symanzik $\beta$-function and $N_0$, $N_{1/2}$, and $N_1$ the quantum numbers or flavours of the corresponding scalar, fermion and $SU(N)$ vector species~\cite{Shifman:1979eb}. This \textit{contact-like} interaction can be alternatively interpreted as a scalaron coupling to the trace of the energy-momentum tensor of the gauge fields, which is now non-vanishing. Taylor expanding the conformal factor in powers of $\phi/M_P$, we get an effective decay width\footnote{If the particles running in the loops are lighter than the scalaron, the anomalous decay $\phi\to A\,A$  can still take place, but the associated rate must be computed directly from the 1-loop diagrams in Fig.~\ref{fig:1loop}. Remarkably, the corrections associated with the intermediate masses appear only as a multiplicative ${\cal O}(1)$ factor. For instance, the decay via bosonic and fermionic loops is given by \cite{Watanabe:2010vy}
\begin{eqnarray}
\Gamma_{\rm anom, SM}^{\rm exact}
=\frac{n_A\, \alpha^2}{1536\, {\pi}^3} \frac{M^3}{M_P^2}\left|\sum _{N_{1/2}}2\,I_{1/2}\left(\frac{M^2}{m^2_{\psi}}\right)+\sum_{N_0} I_0\left(\frac{M^2}{m^2_{\varphi}}\right)\right|^2,
\end{eqnarray}
with 
 $I_{1/2}(M^2/m_{\psi}^2 \to 0)=I_0(M^2/m_{\varphi}^2 \to 0)=1/3$ for heavy internal degrees of freedom and $I_{1/2}(M^2/m_{\psi}^2\to\infty)= 0$ and $I_0(M^2/m_{\varphi}^2\to\infty)\to 2$ for light ones. A similar albeit more involved computation entailing Faddeev-Popov ghosts can be also performed for intermediate gauge bosons \cite{Ellis:1975ap,Shifman:1979eb}. Having in mind these mild corrections, we will stick to Eq.~\eqref{eq:Gamma_a} in our estimates.}
\be\label{eq:Gamma_a}
    \Gamma_{\rm anom, SM} =  \frac{ n_A\, \alpha^2\, b^2 }{1536\,\pi^3} \frac{M^3}{M_P^2}\,,
\ee
with $\alpha=g^2/(4\pi)$ the gauge structure constant. Note that we have implicitly assumed the electroweak symmetry to be unbroken during the heating stage, such that the $n_A$ SM gauge bosons in the final states are massless. 

The dominant contribution to the total scalaron decay width into SM particles,
\be\label{eq:Gamma}
    \Gamma_{\rm SM} = \Gamma_{\rm tree,\, SM} +\Gamma_{\rm anom, \,  SM}\,, 
\ee
depends on the specific matter content. At this point,  we can consider two limiting scenarios: 
\begin{enumerate}

    \item 
    \textit{Tree-level dominated decay.} For a small number of internal particles and provided that the Higgs field is non-conformally coupled to gravity, the tree level decay in Eq.~\eqref{Gamma_varphi} becomes the dominant decay channel, i.e. $\Gamma_{\rm SM}= \Gamma_{{\rm tree, \, SM}}$. Numerically, this corresponds to a decay width
    \be\label{eq:TrhSM}
        \Gamma_{\rm SM}\simeq 2.9 \times 10^{-17}(1 + 6\xi_H)^2 \, M_P \,,
    \ee
    where we have omitted  ${\cal O}(m_H^2/M^2)$ corrections that are important only in the close-to-conformal limit $\xi_H\to -1/6$.
    
    \item 
    \textit{Anomaly dominated decay.} The scalaron decay rate becomes dominated by the anomalous contribution, $\Gamma_{\rm SM}\simeq \Gamma_{\rm anom, \, SM}$, if
    \be\label{N0_estimate}
    \vert b \vert > 20\,\pi \left(\frac{0.1}{\alpha}\right) \left(\frac{16}{n_A}\right)^{1/2} \left(1+6\,\xi_H+2\frac{m_H^2}{M^2}\right)\,.
    \ee
    As shown in Ref.~\cite{Gorbunov:2012ns}, this restriction is easily satisfied in the SM with a conformally coupled Higgs ($\xi_H=-1/6$).\footnote{The close-to-conformal value $\xi_H\approx-1/6$ suppressing direct decays into the Higgs is required at the scale of the inflaton mass. It may be slightly different at the scale of inflation due to the logarithmic running of the non-minimal coupling~\cite{Bezrukov:2009db}.} Alternatively one could consider non-conformally coupled scenarios involving additional SM-charged particles so that $b$ is increased (see e.g~\cite{Takeda:2014qma} and references therein).\footnote{We note, however, that perturbativity might be eventually lost if the number of SM-charged degrees of freedom in a given scenario is too large as quantum corrections scale with $\alpha\cdot  b$. We do not consider such scenarios as a non-perturbative treatment is beyond the scope of this work.} The effective scalaron decay rate can be estimated as
    \be\label{eq:Trh3}
        \Gamma_{\rm SM}\simeq 4.6\, \times\,  10^{-20}\, n_A \, \left(\frac{\alpha}{0.1}\right)^2\left(\frac{b}{10}\right)^2  \, M_P
        \,.
    \ee
    The anomalous contribution due to the SM particle content provides an irreducible contribution to the scalaron decay width and thus implies a rough lower bound on the heating temperature.
\end{enumerate}

The early evolution of the Friedmann-Lema\^itre-Robertson-Walker Universe is determined by the first Friedmann equation, the Klein-Gordon equation for the scalaron field, with its decay rate accounted for by an effective dissipative term~\cite{Kofman:1997yn, Mukaida:2012qn}, and the equation for the radiation energy density, namely
\be\label{eq:bg_ev_t}
	H^2 = \frac{\rho_\phi + \rR}{3\mpl^2}\,, \hspace{8mm}	\ddot \phi + (3H + \Gamma_{\rm SM})\, \dot \phi + V_{,\phi} = 0\,, \hspace{8mm}
	\dot \rho_{R} + 4H\,\rR  = \Gamma_{\rm SM}\, \dot \phi^2\,.
\ee
Here $H$ is the Hubble expansion rate, $\rho_\phi \equiv \dot \phi^2/2 + V$ and $\rR $ denote respectively the scalaron and SM radiation energy densities, and we have approximated the total scalaron decay width by its SM counterpart $\Gamma_\phi\simeq \Gamma_{\rm SM}$, since, by phenomenological reasons, the branching ratio into the DM sector must be small. In terms of the number of $e$-folds $N \equiv \ln a$, these equations read 
\be\label{eq:bg_ev_N}
	H^2=\frac{\rR+V}{3\mpl^2-\frac12{\phi'}^2}\,, 
\hspace{7mm}	\phi'' + \left(3+ \frac{H'}{H} +\frac{\Gamma_{\rm SM}}{H}\right)\phi'+\frac{V_{,\phi}}{H^2}=0\,,
\hspace{7mm}	\rR'+4\,\rR =\Gamma_{\rm SM} H\,{\phi'}^2,
\ee
with the primes denoting derivatives with respect to $N$, and the Hubble damping term 
\be
	\frac{H'}{H} = - \frac{3}{2}\left(1 + \frac{\sum_i P_{i}}{\sum_i \rho_i} \right) = \frac{V + \frac13\rR}{\mpl^2\,H^2} - 3 \in [-3,0]
\ee
following from the continuity equation. 

\begin{figure}[t!]
	\begin{center}
		\includegraphics[height=0.32\textwidth]{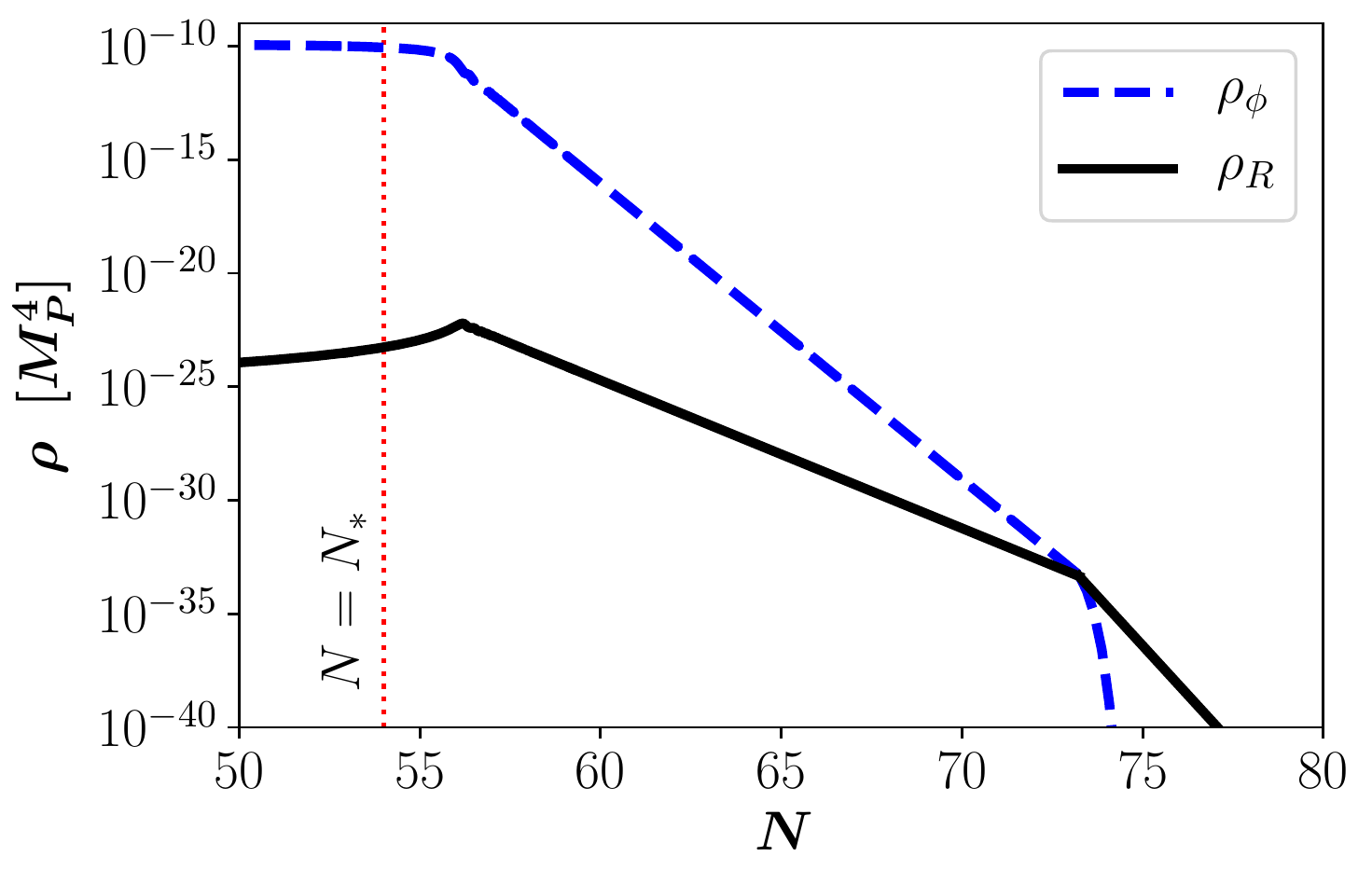}
		\includegraphics[height=0.32\textwidth]{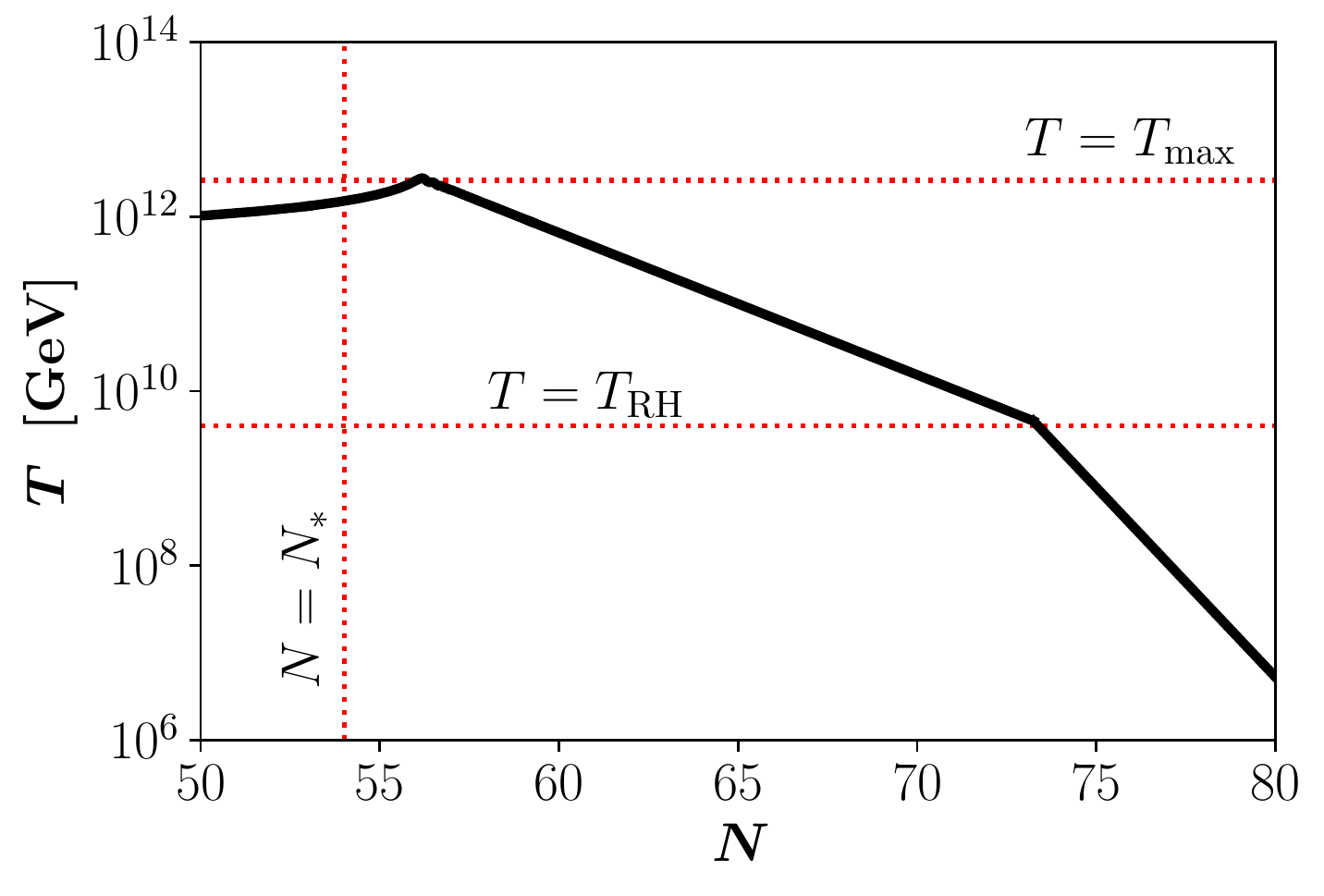}
		\caption{Evolution of the scalaron and the radiation energy densities (left panel), and the radiation temperature (right panel) as a function of the number of $e$-folds. We assumed $N_*=54$ and a SM heating through a minimally coupled Higgs, that is, $\Gamma_\text{SM}=2.9\times10^{-17}~\mpl$.
		}
		\label{fig:Starobinsky2}
	\end{center}
\end{figure}

Slow-roll inflation is followed by a matter-dominated era associated to the coherent oscillations of the scalaron field around its minimum and a radiation-dominated stage following its complete decay into the SM thermal bath. The evolution of the different energy densities during each cosmological epoch can be analytically understood using simple approximations:
\begin{enumerate}
    \item \textit{Inflationary phase.}
As mentioned in Section~\ref{sec:Starobinsky}, the inflationary phase is well described by the slow-roll approximation, where $\rho_{\phi} \simeq V  \simeq 3\mpl^2H^2$ and $3H \dot \phi + V_{,\phi} = 0$. Since the scalar energy density can be taken to be independent from radiation, the equation for the latter can be easily integrated, yielding
\bea\label{eq:rhoSMinfla0}
	\rho_{R} 
&	\simeq \rR(N_0) + \frac{2\mpl}{\sqrt{3}}\int^{N}_{N_0} \td N' \, e^{-4(N-N')}  \sqrt{V} \Gamma_{\rm SM} \,\epsilon_{V}\,,
\eea
with $\epsilon_V \equiv (V'/V)^{2}/2 \mpl^2$ the first (potential) slow-roll parameter. As any initial radiation density $\rR(N_0)$ is diluted during the inflationary stage and the field value is roughly constant within 1/4 $e$-folds, the radiation energy density quickly approaches the stationary value\footnote{This relation is exact when $\sqrt{V}\,\Gamma_{\rm SM} \,\epsilon_{V}$ is constant and $N_0 \to -\infty$.}
\be\label{eq:rhoSMinfla}
    \rho_{R} 
    \simeq \frac{\mpl}{2\sqrt{3}}\sqrt{V}\,\Gamma_{\rm SM}\,\epsilon_V =\frac{\epsilon_V}{6}\,\frac{\Gamma_{\rm \phi}}{H}  \,\rho_\phi  \ll \rho_\phi \,.
\ee
This expression shows explicitly that the radiation energy density is suppressed as compared to the scalaron counterpart, both by the slow-roll parameter $\epsilon_V \ll 1$ and by the small decay rate, $\Gamma_{\rm SM} \ll H$. However, as $\epsilon_V$ grows during inflation, so does $\rR$, reaching its maximal value around the end of inflation, defined by the condition $\epsilon_V = 1$. Assuming instantaneous thermalization of the decay products, the maximal effective radiation temperature of the associated SM plasma can be estimated as
\be\label{eq:Tmax}
	\Tmax = \left[\frac{30}{\pi^2\,\gs} \rR(\Tmax)\right]^{1/4}
	\simeq \left[\frac{5\sqrt{3}}{\pi^2\,\gs} \sqrt{\,V}\mpl\,\Gamma_{\rm SM}\right]^{1/4}\,.
\ee
For a minimally coupled Higgs, this corresponds to 
\be
    \Tmax \simeq 2\times 10^{12}~\text{GeV}\,.
\ee
This is the exact value obtained by numerically solving the evolution equations \eqref{eq:bg_ev_N}. The analytic estimate~\eqref{eq:Tmax} gives a slightly lower value due to the small delay between the end of inflation and the period when the scalaron begins to oscillate. An example of this is depicted in Fig.~\ref{fig:Starobinsky2}.

\item After inflation, the total energy budget of the Universe is dominated by the now oscillating scalaron field,  whose energy density evolves as 
\be\label{rhophi}
	\dot \rho_\phi = -(3 H+\Gamma_{\rm SM}) \dot \phi^2\,.
\ee
Since the frequency of oscillations is determined by the scalaron mass $M$, which significantly exceeds $H$ in this epoch, the scalaron energy density remains approximately constant within a single oscillation period $\tau$. This allows us to replace the square velocity $\dot \phi^2$ in Eq.~\eqref{rhophi} by its time average $\langle \dot \phi^2 \rangle_{\tau}$. Expressing this quantity as a function of $\rho_\phi$~\cite{Turner:1983he},  we get\footnote{In this epoch, the scalaron potential can be approximated by a quadratic expansion around the minimum. By the virial theorem, the time averaged kinetic and potential energies of a $\phi^2$ potential are equal.}
\be
	\langle\dot\phi^2\rangle_{\tau}\simeq \rho_{\phi}\,.
\ee
At early times ($3 H \gg \Gamma_{\rm SM}$), the oscillating scalaron field has an effective equation of state parameter $w_{\phi} \simeq 0$, being its energy density diluted as that of non-relativistic matter, $\rho_{\phi} \propto a^{-3}$.  On the other hand, the density of radiation continues to be determined by the source term $\Gamma_{\rm SM}\, \dot \phi^2$ in Eq.~\eqref{eq:bg_ev_t}.  Plugging $\rR \propto \rho_{\phi}\,\Gamma_{\rm SM}/H$ into the last equation in~\eqref{eq:bg_ev_N} and averaging over oscillations, we obtain
\be\label{eq:rR_osc}
	\rR\simeq  \dfrac{2}{5}\dfrac{\Gamma_{\rm SM}}{H}\, \rho_{\phi} \propto a^{-3/2}\,,
\ee
meaning that the radiation fluid is diluted at half the rate of the scalaron energy density. The temperature of the SM plasma within this period is in the range $\Trh\lesssim T\lesssim\Tmax$, with 
\be
    T_{\rm RH}\simeq \gs^{-1/4} \sqrt{\Gamma_{\rm SM} \, \mpl}
\ee
the standard \textit{reheating temperature}, defined approximately by the moment at which the total scalaron decay width into SM components equals the Hubble rate, $\Gamma_{\rm SM}=3 H$.
For $\gs=106.75$, the tree-level and anomalous scalaron decays in Eqs.~\eqref{eq:TrhSM} and~\eqref{eq:Trh3} yield $T_{\rm RH}\simeq 4.2 \times 10^9 \, {\rm GeV}$ and $T_{\rm RH}\simeq 1.7 \times 10^{8} \left(\frac{\alpha}{0.1}\right) \left(\frac{b}{10}\right) \,{\rm GeV}$ respectively, in agreement with Refs.~\cite{Gorbunov:2010bn,Gorbunov:2012ns}. Note that lower reheating temperatures imply a longer matter-dominated stage between the end of inflation and radiation domination, cf.~Eq.~\eqref{eq:N}. 
\item  After complete heating, the energy density of the scalaron decays exponentially. The SM radiation dominates from there on, scaling in the usual way, $\rR \propto a^{-4}$.
\end{enumerate}

An example of the evolution of the scalaron and radiation energy densities previously described is shown in the left panel of Fig.~\ref{fig:Starobinsky2}. The right panel depicts the corresponding SM temperature.
We have assumed $N_*=54$ and a tree-level dominated decay of the scalaron field into minimally-coupled Higgs bosons, i.e. $\Gamma_\text{SM}=2.9\times10^{-17}~\mpl$. The red dotted lines, corresponding to $T=\Trh$, $T=\Tmax$ and $N_*$, have been added for reference.

Before closing the section, we note that SM particles do not necessarily thermalize instantaneously, and thus the decay products are initially distributed with smaller occupation numbers and harder momenta, $\langle p \rangle \simeq M$ (see e.g. Refs.~\cite{Harigaya:2013vwa, Ellis:2015jpg, Garcia:2018wtq}). In fact, the SM particles are thermalized approximately at a scale~\cite{Harigaya:2013vwa}
\be\label{eq:T_th}
    T_{\rm th} 
    \simeq \alpha^{4/5} M \left( \frac{\Gamma_{\rm SM}\,\mpl^2}{M^3} \right)^{2/5},
\ee
which, for a minimally coupled Higgs and $\alpha \simeq 10^{-2}$, corresponds to $T_{\rm th} \simeq 10^{11}\,\GeV$; an order of magnitude below $\Tmax$. As $\Tmax$ has a milder dependence on $\Gamma_\text{SM}$ than $T_{\rm th}$, the latter can exceed $\Tmax$ if $\Gamma_\text{SM} > 2 \times 10^{-9} \mpl$. Instantaneous thermalization can be safely assumed in this case. If $\Tmax \gtrsim T_{\rm th}$, then $\Tmax$ should be interpreted as an effective quantifier of the maximal radiation density defined through Eq.~\eqref{eq:Tmax}. Non-thermal effects at the moments following inflation can significantly affect UV freeze-in when the DM production cross-section has a sufficiently strong temperature dependence~\cite{Garcia:2018wtq}. We will further develop this point in Section~\ref{sec:nonthermal}.

\section{Dark Matter Production} \label{sec:discussion}

On top of the SM production, any successful cosmological scenario should be able to generate the proper DM abundance. In our particular setting, the evolution of the DM number density $n_\text{DM}$ is given by the Boltzmann equation
\be\label{eq:DMBE0}
	\frac{\td n_\text{DM}}{\td t} + 3H\,n_\text{DM}  = -\sv\left(n_\text{DM}^2-n_\text{eq}^2\right)+2\,\Br\,\Gamma\,\frac{\rho_\phi}{M}\,.
\ee
The first term in the right-hand side of this expression corresponds to the DM production via 2-to-2 scatterings of SM particles, with $\sv$ the corresponding thermally-averaged cross section.
Additionally, $n_\text{eq} = \mathcal{C}_n\,\frac{\zeta(3)}{\pi^2}\,g\,T^3$ is the equilibrium DM number density for ultra-relativistic states in terms of the DM number of degrees of freedom $g$, with  $\mathcal{C}_n=1$ or $3/4$ for bosonic or fermionic DM, respectively.
The second term gives the direct production via the scalaron decay into a couple of DM states.%
\footnote{Thermalization and number-changing processes within the dark sector can have a strong impact on the DM relic abundance. In particular, they can enhance the DM abundance by several orders of magnitude~\cite{Chu:2013jja, Bernal:2015ova, Bernal:2015xba, Bernal:2017mqb, Falkowski:2017uya, Heeba:2018wtf, Mondino:2020lsc, Bernal:2020gzm}.}

Since the number of DM particles is constant after production, it is convenient to express the evolution of the DM density in terms of the co-moving yield $Y\equiv n_\text{DM}/s$, with 
\be
	s(T)\equiv \frac{2\pi^2}{45}\,\gss(T)\,T^3
\ee
the SM entropy density and $\gss(T)$ the effective number of relativistic degrees of freedom contributing to the SM entropy~\cite{Drees:2015exa}. In terms of this quantity, Eq.~\eqref{eq:DMBE0} becomes
\be\label{eq:DMBE}
	\frac{\td Y}{\td T} =\frac{\sv\,s}{H\,T}\left(Y^2-Y_\text{eq}^2\right)-2\,\Br\,\frac{\Gamma}{s\,H\,T}\,\frac{\rho_\phi}{M}\,.
\ee

In the following subsections, we describe in detail the two production mechanisms previously mentioned, namely direct decays and scatterings.

\subsection{Direct Decays}

As we have seen in Section~\ref{sec:Starobinsky}, all type of non Weyl-invariant particles with masses smaller than the scalaron mass can be generated by perturbative gravitational particle production.\footnote{We restrict ourselves to perturbative production of DM below the scalaron mass $M$, acknowledging the potential production of super heavy candidates via non-perturbative processes~\cite{Gorbunov:2012ij}.} This applies also to the DM sector~\cite{Gorbunov:2010bn, Gorbunov:2012ij}. In fact, if DM is mainly produced by the direct 2-body decay of the scalaron field, Eq.~\eqref{eq:DMBE} admits an asymptotic solution $Y(T\ll\Trh)=Y_0$, with
\be 
    Y_0\simeq\frac32\frac{\gs}{\gss}\frac{\Trh}{M}\Br\,.
\ee
In order to reproduce the observed DM energy density $\Omega_\text{DM} h^2\simeq 0.12$~\cite{Aghanim:2018eyx}, the DM yield has been fixed such that 
\be
    m_{\rm DM}\,Y_0 = \Omega_\text{DM} h^2 \, \frac{1}{s_0}\,\frac{\rho_c}{h^2} \simeq 4.3 \times 10^{-10}~\GeV\,,
\ee
with $\rho_c \simeq 1.1 \times 10^{-5} \, h^2$~GeV/cm$^3$ the critical energy density and $s_0\simeq 2.9\times 10^3$~cm$^{-3}$ the present entropy density.
For the scalaron decay not to overclose the Universe, the branching ratio into DM particles must satisfy the bound
\be\label{eq:Br}
	\Br\leq 3\times 10^{-9}\left(\frac{1~\text{TeV}}{m_{\rm DM}}\right)\left(\frac{54}{N_*}\right)\left(\frac{2.8\times 10^9~\text{GeV}}{\Trh}\right),
\ee
with the equality associated to the case in which the scalaron produces the whole DM relic abundance.
Let us emphasize again that the scalaron field must decay mainly into SM particles ($\Br\ll 1$), and thus Eq.~\eqref{eq:Br} implies that DM has to be heavier than $m_\text{DM}\gg 3$~keV.
Depending on the DM spin, we can distinguish three scenarios:
\begin{enumerate}
    \item \textit{Spin-0 DM.} The scalaron efficiently decays into scalar states through derivative interactions. In particular, if heating proceeds through a minimally coupled Higgs, then the branching fraction for a minimally-coupled real scalar DM candidate is $\sim 1/5$, cf. Eqs.~\eqref{Gamma_varphi} and~\eqref{eq:Gamma}, clearly overshooting the observed relic abundance for cold DM candidates. This branching ratio can be however tuned in the presence of a non-minimal coupling $\xi_\text{DM}$ between the scalar DM component and gravity. The observed DM abundance is then obtained for
    \be
        1+6\,\xi_\text{DM} \simeq 10^{-4}\sqrt{\frac{1\,\TeV}{m_{\rm DM}}}\,,
    \ee
    with a mild dependence on the Higgs non-minimal coupling. Figure~\ref{fig:ScalarDM-directdecay} shows the non-minimal coupling $\xi_{\rm DM}$ required for the production of scalar DM via the decay of the scalaron field. In particular we notice that this mechanism is only compatible with DM masses below $m_{\rm DM} \lesssim 10^{10}$~GeV for non-minimal DM couplings and $m_{\rm DM}\simeq 10$~keV if DM is minimally coupled. These results agree with those in Refs.~\cite{Gorbunov:2010bn,Gorbunov:2012ij}.
    \begin{figure}[t!]
	    \begin{center}
		    \includegraphics[height=0.33\textwidth]{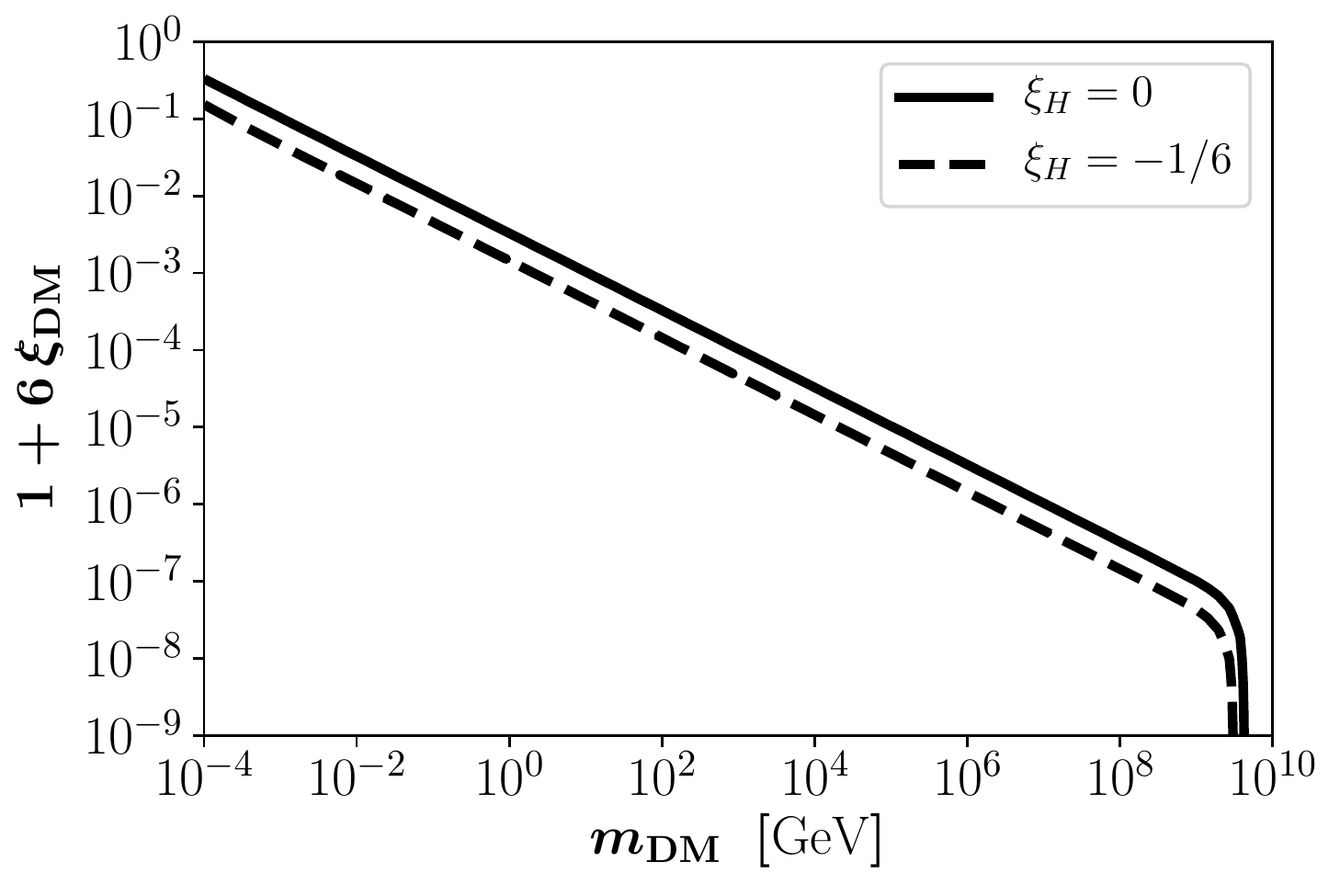}
		    \caption{Non-minimal coupling required for the production of scalar DM via the decay of the scalaron for $\xi_H=0$ (solid line) and $\xi_H=-1/6$ (dashed line).		    }
		    \label{fig:ScalarDM-directdecay}
	    \end{center}
    \end{figure}
    \item \textit{Spin-1/2 DM.} 
    A fermionic DM candidate with explicit mass term will couple to the scalaron when moving to the scalaron frame. The associated decay width is given by 
    \be\label{eq:Gamma12}
        \Gamma_{1/2} =\frac{1}{48\pi}\frac{m_{\rm DM}^2}{M^2}\frac{M^3}{\mpl^2}\left(1-\frac{4\,m_{\rm DM}^2}{M^2}\right)^{3/2}. 
    \ee
 The direct decay of the scalaron in this case can only produce DM particles of mass $m_{\rm DM}\simeq 10^7$~GeV, with a very small dependence on the Higgs non-minimal coupling. This result agrees with that in Ref.~\cite{Gorbunov:2010bn}. 
    \item \textit{Spin-1 DM.} The scalaron can decay into vector DM via anomalous processes generated by $b$ fields coupled to the DM charge. Recovering the observed DM abundance in this case requires (cf.~Eqs.~\eqref{eq:Gamma} and~\eqref{eq:Trh3})
    \be
        \alpha\cdot b\simeq 3\,\sqrt{\frac{100~\MeV}{m_{\rm DM}}}\,,
    \ee
with a very small dependence on the Higgs non-minimal coupling to gravity. 
\end{enumerate}

\subsection{UV Freeze-in}
\label{sec:UV_FI}

In the previous section we have seen that the direct production of DM out of the scalaron field is either not phenomenologically viable, restricted to specific mass values, or associated to a close-to-conformal scalar field. Another production channels are therefore generically required. Here we study an \textit{alternative mechanism working independently of the DM spin} (although we will assume a conformal coupling for the scalar DM hereafter) and taking place in a time window between the creation and thermalization of the SM plasma ($t_{\rm SM}\sim \Gamma_{\rm SM}^{-1} $) and the time at which the direct DM production out of the scalaron field would become relevant ($t_\text{direct}\sim (M^2/m_{\rm DM}^2)\, t_{\rm SM}$).

If the interaction rates between the visible and the dark sectors were never strong enough, the observed DM relic abundance could have been produced in the early Universe by non-thermal processes. 
One possibility for having small DM production rates is to consider non-renormalizable operators $\bigO_{D}$ with mass dimension $D>4$.
This corresponds to the so-called UV freeze-in mechanism, where the thermally-averaged DM annihilation cross section $\sv$ in Eq.~\eqref{eq:DMBE0} has typically a strong temperature dependence
\be\label{eq:sv}
	\sv=\frac{T^n}{\Lambda^{n+2}}\,,
\ee
with $n\geq 0$ and $\Lambda$ the cutoff scale of the effective field theory, potentially interpreted as a proxy of the mediator mass connecting the dark and visible sectors.\footnote{Non-conformally coupled scalars can interact via a scalaron-mediated $s$-channel. Due to the derivative couplings, the corresponding cross section can scale strongly with temperature. For example, the production of scalar DM $S$ via Higgs annihilation is given by
\be
   \sigma_{HH \to SS^*} = \frac{(1+6\xi_H)^2 (1+6\xi_\text{DM})^2}{576\pi\,\mpl^4\,M^4}\,s^3 + \bigO(s^2)\,,
\ee
for $m_{H}^2 \ll s \ll M^2 $ and $s$ the center of mass energy squared. However, this channel is strongly suppressed by high powers of $M$ and $M_P$ and therefore subdominant with respect to the production via scalaron decays.} This cross section is generated by a non-renormalizable operator with mass dimension $D=5+n/2$, for $n$ even and non-negative.
Additionally, for this effective operator description to be valid, $\Lambda$ must be the highest scale in the calculation, and we assume throughout the hierarchy $m_{\rm DM}\ll \Trh<\Tmax\ll\Lambda$.

In the case where the DM abundance is mainly produced in the early Universe by annihilations of SM particles, Eq.~\eqref{eq:DMBE} can be rewritten as
\be\label{eq:Yield}
	\frac{dY}{dT} \simeq \frac{\sv\,s}{H\,T}\left(Y^2-Y_\text{eq}^2\right)\simeq-\frac{\sv\,s}{H\,T}Y_\text{eq}^2\,,
\ee
where we have used the fact that DM never reaches the equilibrium distribution in freeze-in production, i.e. $Y\ll Y_\text{eq}$. At $T\ll\Trh$, this equation admits the asymptotic solution\footnote{When $n=6$, the term in the square brackets is $\ln(\Tmax/\Trh)$ .}~\cite{Garcia:2017tuj, Bernal:2019mhf}
\bea\label{eq:DMYield}
	Y_0
&    \simeq \frac{45\,\zeta(3)^2\,\mathcal{C}_n^2}{\pi^7}\sqrt{\frac{10}{\gs}}\frac{g^2}{\gss}
    \frac{\mpl\,\Trh^{n+1}}{\Lambda^{n+2}} \left\{\frac{3}{2(n+1)} + 4\left[\frac{\left(\Tmax/\Trh\right)^{n-6}-1}{n-6}\right] \right\}.
\eea
Few comments are in order here: $i)$ the first term in this equation arises from the production below $\Trh$, where the Hubble expansion rate is driven by the SM energy density. This corresponds to the usual assumption of a sudden decay of the scalaron field. $ii)$ However, as we have seen in Section~\ref{sec:reheat}, the decay of the scalaron is rather a continuous process. In fact, during the period $\Tmax>T>\Trh$, the energy density of the Universe is still dominated by the oscillating scalaron component, which displays an effective equation-of-state parameter  $w\simeq0$ and leads to a scaling $H\propto a^{-\frac32}$ for the Hubble rate.
Additionally, as the scalaron decays into SM radiation, the SM entropy is not conserved and the photon temperature scales as $T\propto a^{-\frac38}$. The DM production previous to complete heating is estimated by the second term in Eq.~\eqref{eq:DMYield}. This term dominates if the DM annihilation cross section has a strong temperature dependence, and can appear as a logarithmic boost $\ln(\Tmax/\Trh)$ for $n=6$, or power-law boost $(\Tmax/\Trh)^{n-6}$ for $n>6$, relative to the abundance in the instant decay approximation~\cite{Garcia:2017tuj, Bernal:2019mhf}. $iii)$ Finally, we note that the inflaton decay products typically thermalize on a much shorter timescale than $\Gamma_\phi^{-1}$, so it is appropriate to assume that the thermal bath existed before heating~\cite{Davidson:2000er, Harigaya:2013vwa, Mukaida:2015ria, Ellis:2015jpg, Harigaya:2019tzu}. Since the DM abundance in Eq.~\eqref{eq:DMYield} is mildly dependent on $\Tmax$ for $n\leq 6$, the thermal abundance is mostly insensitive to potential physics around $\Tmax$. However, as thermalization can take place at temperatures slightly lower than $\Tmax$, the non-thermal contribution must be studied separately.

\begin{figure}[t!]
	\begin{center}
		\includegraphics[height=0.5\textwidth]{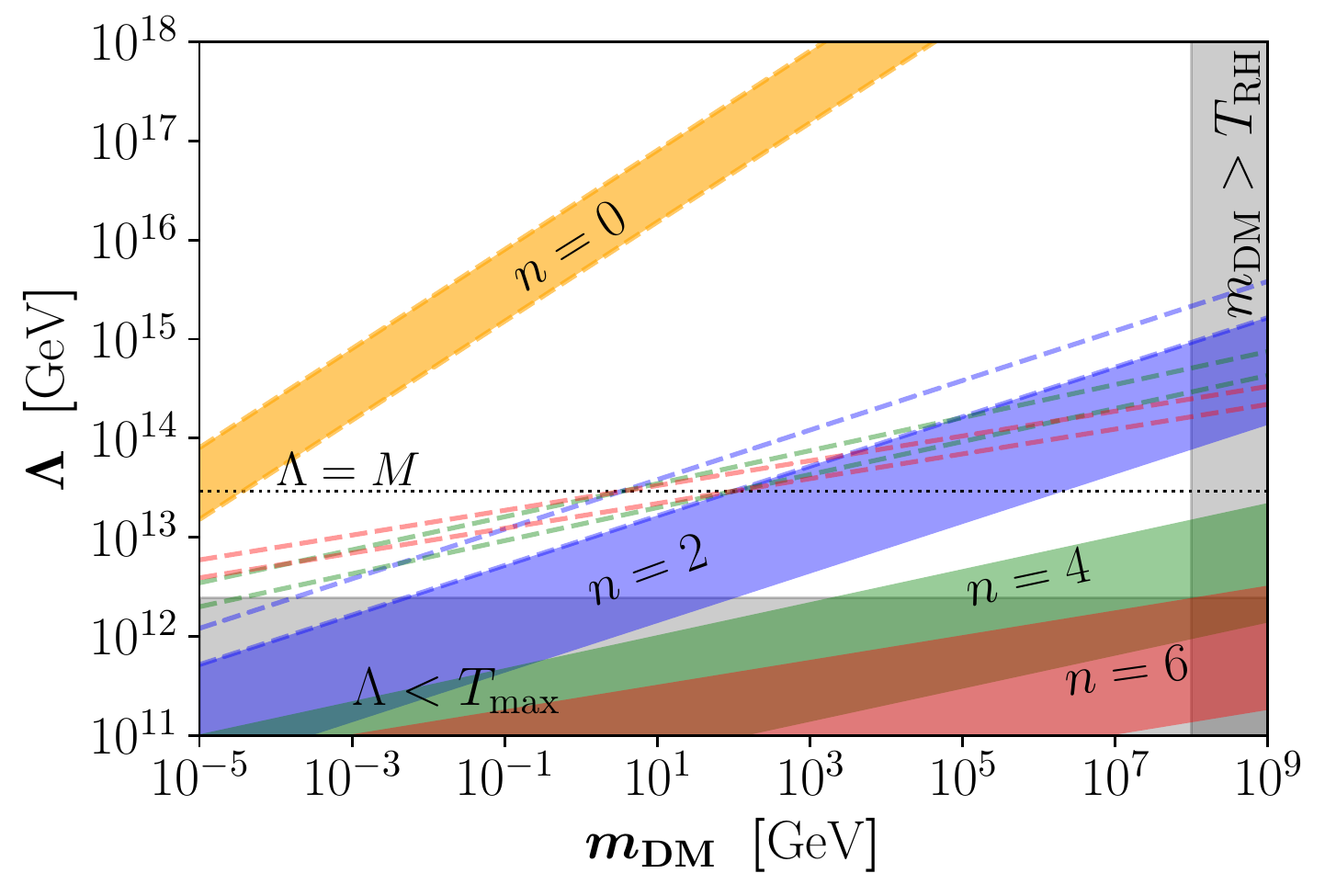}
		\caption{Parameter space reproducing the observed DM relic abundance for different temperature exponents $n$, via 2-to-2 scatterings only (colored bands) or scatterings and scalaron decays (dashed lines).
		The thickness of the bands is due to the limiting cases in 
		Eqs.~\eqref{eq:TrhSM} corresponding to reheating into the minimally coupled Higgs (upper bounds) and~\eqref{eq:Trh3} corresponding to reheating via anomaly induced scalaron decays (lower bounds).
		The gray areas correspond to regions $\Lambda<\Tmax$ and $m_{\rm DM}>\Trh$ outside the expected limit of validity of the effective field theory. Reheating through the minimally coupled Higgs is assumed in this figure.
		}
		\label{fig:StarDM}
	\end{center}
\end{figure}

The colored bands in Fig.~\ref{fig:StarDM} show the parameter space reproducing the observed DM relic abundance  via the UV freeze-in mechanism for different values of $n$. The thickness of the bands is related to the uncertainty of the scalaron width, which could be either dominated by tree-level (upper bounds) or anomaly-induced (lower bounds) decays. The gray areas stand for the regions $\Lambda<\Tmax$ and $m>\Trh$ where our effective field theory approach ceases to be trustable. We find that $n=6$ case is at odds with the effective field theory approach.

\subsection{Non-thermal Effects}
\label{sec:nonthermal}
The non-thermal mode for freeze-in production of DM is active when $\Tmax \gtrsim T \gtrsim T_{\rm th}$. Due to the linearity of the Boltzmann equations for freeze-in, we can split the DM abundance to a thermally and a non-thermally produced component $n_{\rm DM, NT}$ satisfying 
\be\label{eq:DMBE_NT}
    \frac{\td n_{\rm DM, NT}}{\td t} + 3H n_{\rm DM, NT} \simeq \sv_{\rm NT} \, n_{\rm SM, NT}^2\,.
\ee
Assuming that the cross-section responsible for DM production scales as $\sigma \propto E^n$, the cross-section $\sv_\text{NT}$ averaged over the non-thermal component and relevant when $\Tmax > T > T_{\rm th}$, is approximately given by~\cite{Garcia:2018wtq}
\be\label{eq:sv_NT}
    \sv_\text{NT} 
    \simeq \sv \left( \frac{M}{13\, T} \right)^{n}
    = \frac{1}{\Lambda^{2}}\left( \frac{M}{13\, \Lambda} \right)^{n},
\ee
where the numerical factor 13 was obtained by fitting the more precise expression given in Ref.~\cite{Garcia:2018wtq}. In the minimally coupled Higgs scenario, $\sv_\text{NT} \gtrsim \sv$ for $T \lesssim \Tmax$, and the DM production from the non-thermal component is thus always enhanced. Note, however, that Eq.~\eqref{eq:sv_NT} assumes the dominant production channel to be the same for both the non-thermal and thermal components of SM radiation. For example, if the scalaron decays dominantly into Higgs particles but the thermal production \eqref{eq:sv} receives a negligible contribution from Higgs annihilation, then the enhancement implied by Eq.~\eqref{eq:sv_NT} is not realized. In conclusion, when non-thermal effects are relevant, one may have to specify the dominant channels of DM production, and work out the particularities of visible sector thermalization.

The number density of scalarons satisfies $n_{\phi} \simeq n_{\phi,0}\,(a_0/a)^3\,e^{-v}$, where $v \simeq \Gamma_{\rm SM} (t - t_{0}) \simeq 3\Gamma_{\rm SM}/(2H)$ and $t_{0}$ denotes the point where the scalaron starts oscillating. As thermalization takes place before reheating, we can safely assume $v \ll 1$ or, equivalently, $H \gg \Gamma_\phi$. In the absence of number-changing interactions and assuming 2-body decays of the inflaton, it holds that $2\, n_{\phi} + n_{\rm SM} \propto a^{-3}$, and therefore we have  $n_{\rm SM} \simeq 2 n_{\phi}\,v$. Following Refs.~\cite{Harigaya:2013vwa,Harigaya:2014waa,Mukaida:2015ria,Harigaya:2019tzu}, the effect of depletion of the non-thermal SM component and number density of non-thermal SM particles can be estimated as
\be\label{eq:n_NT}
    n_{\rm SM, NT} 
    \simeq 2 n_{\phi}\,v \left[ 1 - \left(\frac{2 k_{\rm split}}{M}\right)^{3/2} \right],
\ee
where
\be
    k_{\rm split} 
    = \frac{\alpha^{16}\mpl^6\,v^5}{2 M^3 \Gamma_{\rm SM}^2} 
    = \frac{M}{2}\left(\frac45\frac{T_{\rm th}^4}{\rho_{R}}\right)^5 
\ee
is the splitting momentum below which a particle can deposit a significant fraction of its energy and participate in the thermal plasma. Noting that $\rho_{R} = (2/5) M\,n_{\phi}\,v \propto H$, the Boltzmann equation \eqref{eq:DMBE_NT} can be easily integrated in terms of the number of $e$-folds, yielding the non-thermal DM number density at the time of thermalization
$
    n_{\rm DM, NT}|_{T = T_{\rm th}} 
    \simeq 2 \sv n_{\phi}^2 v^2/H|_{T = T_{\rm th}} \,.
$
This expression holds when the beginning of the scalaron oscillations and thermalization are separated by $N \gtrsim 1$ $e$-folds. Otherwise, $n_{\rm SM, NT}$ is suppressed by a factor proportional to $N^3$. After thermalization, the comoving number density of this DM component is conserved. Thus the resulting DM yield is
\bea
    Y_{\rm NT}
&    = \frac{n_{\rm DM, NT}(T_{\rm th})}{s_{\rm RH}} \left( \frac{a_{\rm th}}{a_{\rm RH}} \right)^{3}
    \simeq \frac{g^2}{\gss}\frac{\Trh^{5} \sv_{\rm NT}}{H_{\rm th} M^2} \\
&    \simeq \mathcal{O}(10^{-3}) \left(\frac{\alpha}{10^{-2}}\right)^{-16/5} \left(\frac{\Trh}{2.4 \times 10^{12}~\text{GeV}} \right)^{14/5-n} Y_0\,,
\eea
where the second line shows the explicit comparison with the thermal yield \eqref{eq:DMYield} when $n\leq 6$. When $\Trh \ll 2.4 \times 10^{12}$~GeV, the last term in the brackets can dominate when $n\gtrsim 3$, as pointed out in Ref.~\cite{Garcia:2018wtq}. However, a direct numerical comparison of $Y_{\rm NT}$ with $Y_{0}$ in the minimally coupled Higgs scenario shows that the non-thermal contribution is subdominant when $n < 4$ and comparable when $n=4$. On the other hand, the $n \geq 6$, the non-thermal DM yield exceeds the thermal one by a factor of $\mathcal{O}(10^{5})$.

An increase of the yield allows weaker interactions and thus higher values for $\Lambda$. Non-thermal effects will thus shift the $n=6$ band in Fig.~\ref{fig:StarDM} up by about a factor of 10. However, in general, it will not be sufficient to compete with the production via induced scalaron decays. All in all, we find that the non-thermal effects can be safely neglected if $n \leq 2$, while for larger $n$, the DM abundance estimate cannot a priori omit the, potentially model dependent, contributions from non-thermal effects and induced scalaron decays.

\subsection{UV Sector Induced Scalaron Decay}

The operator $\bigO_{D}$ responsible for the scattering cross section~\eqref{eq:sv} is not Weyl invariant for $D \geq 4$. Thus, the conformal transformation~\eqref{eq:Omega} will generically generate a scalaron-frame coupling of the form\footnote{We remark that, on top of the simple rescaling assumed in this expression, additive contributions might be generated by the Weyl rescaling if the effective operator contains also derivatives.} 
\be \label{On}
    \Omega^{D-4}\,\frac{\bigO_{D}}{\Lambda^{D-4}} 
    \simeq \frac{\bigO_{D}}{\Lambda^{D-4}}   + \frac{D-4}{\sqrt{6}} \frac{\phi}{\mpl}
    \frac{\bigO_{D}}{\Lambda^{D-4}} + \ldots
\ee
The last term in the right hand side of this expression contributes to the direct scalaron decay into the dark sector, inducing a decay width of order
\be\label{eq:Gamma_DM_O}
    \Gamma_{\bigO} \simeq 10^{-6} \, \left( \frac{M}{\Lambda} \right)^{2(D-4)}\,\Gamma_* \,,
\ee
where $\Gamma_* \equiv M^3/(24\pi\mpl^2)$ corresponds to the minimally coupled Higgs case given in Eq.~\eqref{Gamma_varphi}. To have a consistent UV freeze-in scenario, this channel should not dominate the DM production.  Compared to the scattering cross-section~\eqref{eq:sv}, the decay width~\eqref{eq:Gamma_DM_O} is further suppressed by phase-space factors. More precisely, in deriving this rough estimate, we assumed UV freeze-in via 2-to-2 scattering. In the scalaron frame, this implies a scalaron decay into two dark and two SM particles, yielding a phase-space suppression of about $(3072 \pi^4)^{-1} \simeq 3\times 10^{-6}$ when compared to the phase space of the scattering process.

In order to avoid the production of states living at the scale $\Lambda$ (e.g. heavy mediators) out of the perturbative scalaron decay, we must require $\Lambda \gtrsim M$.  This implies that the branching ratio for the decay \eqref{eq:Gamma_DM_O} is suppressed at least by a factor of $10^{-6}$, assuming heating to proceed through a minimally coupled Higgs. In this case, UV freeze-in with $n=6$ is disfavoured, as shown in Fig.~\ref{fig:StarDM}.

The induced scalaron decay contributes to the DM abundance and may even dominate DM production. The dashed lines in Fig.~\ref{fig:StarDM} indicate the parameter space that reproduces the DM abundance, taking into account both 2-to-2 scatterings and the production via the decay of the scalaron, estimated using Eq.~\eqref{eq:Gamma_DM_O}. In the case $n=0$, the DM production is completely dominated by the scattering, while for $n=2$ the two production mechanisms can be of the same order of magnitude. When $n > 2$ the decays, given by Eq.~\eqref{eq:Gamma_DM_O}, are expected to dominate. We stress, however, that Eq.~\eqref{eq:Gamma_DM_O} is only a rough order of magnitude estimate. The interplay between DM production via induced decays and scatterings must be worked out in the specific model under consideration. For example, UV freeze-in in the $n>2$ case may still dominate if the induced scalaron coupling is strongly suppressed for on-shell scalarons. 

\begin{figure}[t!]
	\begin{center}
$\vcenter{\hbox{		\includegraphics[width=0.3\textwidth]{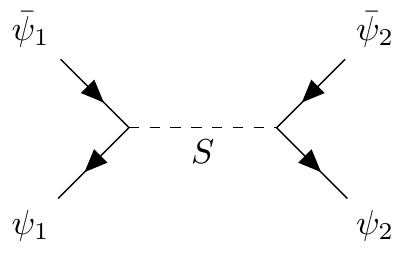}
}}$
		\hspace{2cm}
$\vcenter{\hbox{		\includegraphics[width=0.3\textwidth]{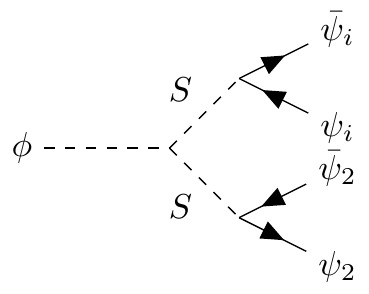}
}}$
\caption{The competing interactions for the production of the $\psi_2$ particle in our toy model: $\psi_1$ annihilation into $\psi_2$ responsible for UV freeze-in (left panel) and the induced 4-body decay of the scalaron (right panel).}\label{fig:4body}
	\end{center}
\end{figure}

The specific realization of the UV freeze-in depends on unknown UV physics beyond the theory cutoff. In order to obtain a clearer picture of the relation between scatterings and direct decays, it is instructive to consider a simple but explicit toy model for UV freeze-in. To this purpose, we choose a heavy scalar $S$ with mass $m_S \gtrsim M$ coupled to two fermions $\psi_{1}$, $\psi_{2}$ representing the visible and DM sectors, respectively. We assume trilinear interactions $y_i\,S\,\bar \psi_{i}\,\psi_{i}$, with $y_i$ denoting Yukawa couplings, cf. left panel of Fig.~\ref{fig:4body}. At energies $s \ll M^2$, the two sectors are connected via a cross section
\be
    \sigma_{11 \to 22} 
    \simeq \frac{y_{1}^2y_{2}^2}{16\pi m_S^4} s\,, 
\ee
leading to a thermally-averaged cross section
\be
    \langle \sigma_{11 \to 22} v \rangle \simeq 0.4 \frac{T^2}{\Lambda^4},
    \qquad \qquad \mbox{with} \qquad \qquad
    \Lambda \equiv \frac{m_S}{\sqrt{y_1\,y_2}}\,.
\ee

In the scalaron frame, the Weyl rescaling couples the scalaron field to the heavy scalar $S$ via a $\phi SS$ interaction as that following from Eq.~\eqref{SM_EF} at small $\phi$ values. This produces a tree-level four-particle decay $\phi \to \bar \psi_{1} \psi_{1} \bar \psi_{2} \psi_{2}$ through a pair of off-shell $S$ mediators, cf. right panel of  Fig.~\ref{fig:4body}.\footnote{We remark that the 2-body decay $\phi \to \bar \psi_{i} \psi_{i}$ is induced at one loop with a decay width proportional to the fermion mass. Thus, the loop-order decay is subleading as compared to the tree level decay \eqref{eq:Gamma12} of massive fermions and can be neglected in this order of magnitude analysis.} Assuming the fermion masses to be much smaller than the scalaron mass, $m_{1,2} \ll M$, the corresponding decay widths at leading order are given by 
\bea\label{eq:Gamma_DM_toy}
    \Gamma_{\phi \to 1122} 
&    \simeq \frac{\Gamma_*}{3072 \pi^4}\frac{M^4 y_1^2 y_2^2}{15  m_S^4} 
    \simeq 2 \times 10^{-7} \, \left(\frac{M}{\Lambda}\right)^4 \,  \Gamma_*\,,
    \\
    \Gamma_{\phi \to iiii} %
&    \simeq  \frac{\Gamma_*}{3072 \pi^4} \frac{M^4 y_i^4}{40  m_S^4} 
    \simeq 9 \times 10^{-8} \,\left(\frac{M}{\Lambda}\right)^4 \left(\frac{y_i^2}{y_{_{i'}}^2}\right)\, \Gamma_*\,,
\eea
with the numerator of the first fraction corresponding to the minimally coupled Higgs case in Eq.~\eqref{Gamma_varphi} and the denominator accounting for the suppression due to the 4-body phase space, such that the decay widths have the same form as \eqref{eq:Gamma_DM_O}. We neglected the contribution from the derivative couplings between the Higgs and the scalaron since the $\phi SS$ interaction is generally dominated by the mass term for $m_S \gtrsim M$.\footnote{The non-minimal coupling becomes relevant when $|\xi_S| \gtrsim m_S^2/M^2$. Moreover, in case $\xi_S \simeq -1/6 - m_S^2/(2M^2)$, the Weyl-induced trilinear couplings of the on-shell $\phi$, and therefore the decay of $\phi$, are suppressed.} We remark that the prefactor in Eq.~\eqref{eq:Gamma_DM_toy} is about an order of magnitude smaller than in general estimate in Eq.~\eqref{eq:Gamma_DM_O}. This is sufficient to suppress the decay into DM so that, in the toy model, the DM  production due to UV freeze-in can compete with the production via direct decays.

In all, we see that all mediator-induced scalaron decay widths in our toy model are consistent with the above dimensional analysis. We further see that, due to the phase-space suppression, the induced decay widths are several orders of magnitude smaller than $\Gamma_*$ as long as $\Lambda > M$ and the interactions between the mediator and the visible and dark sectors are of relatively similar strengths; in the toy model, $\min(y_{1}/y_{2},\,y_{2}/y_{1}) \lesssim 10^{-4}$.

\section{Conclusions and Discussion} \label{sec:end}

Despite decades of experimental efforts, non-gravitational interactions between the dark and the visible sector have not been found. This has turned the attention away from the standard WIMP paradigm, motivating the quest of alternative scenarios where DM is produced by feeble interactions. A natural possibility that has recently received considerably attention is the use of non-renormalizable operators involving large cutoff scales typically associated with the mass of a heavy mediator. Given the extreme temperature dependence of these operators, it is natural to embed them in specific heating models where the maximum temperature of the SM plasma can be accurately computed. One remarkable scenario in this regard is the Starobinsky model of inflation, where the interactions among the inflaton field and the SM constituents are completely dictated by conformal symmetry.

We considered the generation of the observed DM abundance in the Starobinsky model of inflation, showing that its perturbative production from the direct decay of the inflaton field is only viable for a very narrow spectrum of close-to-conformal scalar fields and spinors of mass $\sim 10^7$~GeV.
This spectrum can be, however, significantly broadened in the presence of non-renormalizable interactions between the dark and the visible sectors. In particular, we showed that UV freeze-in can efficiently generate the right DM abundance for a large range of masses spanning from the keV to the PeV scale and arbitrary spin, without significantly altering the heating dynamics. However, the higher dimensional operators responsible for UV freeze-in will also generate additional decay channels for the inflaton field and we find that, when the cross section scales as $\sv\propto T^n$ with $n \geq 4$ they cannot be a priori neglected. Similarly, we further found that, when $n \geq 4$, the instantaneous thermalization approximation may fail as, in this case, the non-thermal contribution to the DM yield will likely dominate.

Although the scenario presented here assumes a trivial dark sector dynamics, it is conceivable that it could display sufficiently strong interactions leading to self-thermalization. This could have a strong impact on the DM abundance, opening the possibility of having a dark freeze-out in the dark sector~\cite{Chu:2013jja, Bernal:2015ova, Bernal:2015xba, Bernal:2017mqb, Falkowski:2017uya, Heeba:2018wtf, Mondino:2020lsc}. 
In particular, if the dark sector could thermalize within itself, its entropy would be conserved instead of its number density, leading to an additional enhancement of the DM abundance~\cite{Bernal:2020gzm}.

\section*{Acknowledgments}

The authors thank Fedor Bezrukov, Juan Garc\'ia-Bellido and Tommi Markkanen for useful discussions. This work was supported by the Estonian Research Council grants PRG803, MOBTTP135, and MOBTT5 and by the EU through the European Regional Development Fund CoE program TK133 ``The Dark Side of the Universe”.
Additionally, this project has received funding from the European Union's Horizon 2020 research and innovation program under the Marie Sk\l{}odowska-Curie grant agreements 674896 and 690575, and from Universidad Antonio Nariño grants 2018204, 2019101 and 2019248. NB is partially supported by Spanish MINECO under Grant FPA2017-84543-P. JR acknowledges the support of the Fundação para a Ciência e a Tecnologia (Portugal) through the CEEC-IND programme, grant CEECIND/01091/2018.

\bibliographystyle{JHEP}

\bibliography{biblio}

\end{document}